\documentclass{Interspeech2024}
\usepackage{amsmath}
\usepackage{multirow,makecell}
\usepackage{tabularray}
\DeclareMathOperator*{\argmax}{arg\,max}

\interspeechcameraready

\title{Multi-Channel Multi-Speaker ASR Using Target Speaker's Solo Segment}

\name[affiliation={1}]{Yiwen}{Shao}
\name[affiliation={2*}]{Shi-Xiong}{Zhang}
\name[affiliation={2}]{Yong}{Xu}
\name[affiliation={2}]{Meng}{Yu}
\name[affiliation={2}]{Dong}{Yu}
\name[affiliation={3}]{Daniel}{Povey}
\name[affiliation={1}]{Sanjeev}{Khudanpur}

\address{
  $^1$Center for Language and Speech Processing, Johns Hopkins University, Baltimore, MD, USA\\
  $^2$Tencent AI Lab, Bellevue, WA, USA\\
  $^3$Xiaomi Corp., Beijing, China}

\email{yshao18@jhu.edu}

\begin{document}

\maketitle
\renewcommand*{\thefootnote}{\fnsymbol{footnote}}
\footnote{This work was done while Shi-Xiong was at Tencent AI Lab, USA. }
\begin{abstract}

In the field of multi-channel, multi-speaker Automatic Speech Recognition (ASR), the task of discerning and accurately transcribing a target speaker's speech within background noise remains a formidable challenge. Traditional approaches often rely on microphone array configurations and the information of the target speaker's location or voiceprint. This study introduces the Solo Spatial Feature (Solo-SF), an innovative method that utilizes a target speaker's isolated speech segment to enhance ASR performance, thereby circumventing the need for conventional inputs like microphone array layouts. We explore effective strategies for selecting optimal solo segments, a crucial aspect for Solo-SF's success. Through evaluations conducted on the AliMeeting dataset and AISHELL-1 simulations, Solo-SF demonstrates superior performance over existing techniques, significantly lowering Character Error Rates (CER) in various test conditions. Our findings highlight Solo-SF's potential as an effective solution for addressing the complexities of multi-channel, multi-speaker ASR tasks.

\end{abstract}

\section{Introduction}

\label{sec:intro}
Recent advancements in speech processing techniques and deep learning have led to substantial improvements across various automatic speech recognition (ASR) benchmarks \cite{hinton2012deep, wang2019espresso, shao2020pychain, zhang2020pushing}. However, accurately recognizing multi-channel, multi-speaker overlapped speech remains challenging, largely due to interfering speakers and background noise \cite{haeb2020far, masuyama2023end}. In this complex ASR landscape, the utilization of high-quality, discriminative input features beyond traditional spectral features is crucial for isolating target speech from mixtures. Spatial features, leveraging the phase difference across microphone channels caused by the distinct locations of audio sources, have gained considerable attention \cite{chen2018multi}. These features form the foundation of many state-of-the-art speech separation \cite{gu2020multi,zhang2021adl} and recognition systems \cite{yu2021audio,shao2022multi}.

Extending this line of inquiry, Shao et al. introduced a novel perspective by utilizing the room impulse response (RIR) from the target speaker to the microphone array, enhancing spatial feature extraction to include RIR-based spatial features (RIR-SF) \cite{shao2023rir}. Although RIR-SF approaches an ideal solution with accessible ground truth RIR, the practical acquisition of accurate RIR poses significant challenges, affecting the robustness of RIR-SF in real-world applications.

Addressing these limitations, our study proposes leveraging a short solo segment from the target speaker as an innovative proxy for the actual RIR. This Solo-SF method, through convolution with the overlapped speech signal, aims to overcome the challenges of direct RIR usage by harnessing the unique vocal characteristics of the target speaker. Notably, our approach eliminates the dependency on microphone topology and vision-based positional inputs, facilitating its future application as a universal encoder \cite{huang2023unix} for diverse multi-channel data setups. Furthermore, we delve into strategies for selecting optimal solo segments and assess our method's efficacy on both the simulated AISHELL-1 \cite{bu2017aishell} dataset and the real-world AliMeeting dataset \cite{yu2022m2met}, demonstrating significant performance enhancements.

\vspace{-0.2cm}
\section{Preliminary: Spatial Feature}
\vspace{-0.1cm}
Spatial Feature (SF) or Angle Feature (AF), originally introduced by Chen et al.~\cite{chen2018multi}, are utilized to underscore the prominence of the target sound source within multi-speaker Time-Frequency (T-F) bins. Unlike most spectral features that depend on the magnitude of the complex Short-Time Fourier Transform (STFT) coefficients \(Y \in \mathbb{C}^{T \times F \times M}\), where \(T\), \(F\), and \(M\) represent the time, frequency, and channel dimensions respectively, SF is specifically designed to be phase-sensitive. Spatial feature, denoted as \(\text{SF} \in \mathbb{R}^{T \times F}\), leverages the phase differences across channels, attributed to the disparate spatial locations of sound sources. This phase-sensitive design enables SF to effectively differentiate between sources based on their unique spatial characteristics.

\subsection{3D spatial feature}
Given a pair of microphones, denoted as \(p = (m_1, m_2)\), within a microphone array, and the multi-channel STFT of the input speech signal, represented as \(Y \in \mathbb{C}^{T \times F \times M}\), two types of interchannel phase differences are defined: the target-independent interchannel phase difference (IPD) and the target-dependent interchannel phase difference (TPD) as follows:

\begin{equation}
     \text{IPD}_{t,f,(p)} = \angle Y_{t,f, m_1} - \angle {Y}_{t,f,m_2}
    \label{ipd}
\end{equation}
\begin{equation}
\begin{split}
   & \text{TPD}_{t,f, (p)} (\theta_a, \theta_e, d_o) = \frac{2\pi f}{c(F-1)} \cdot f_s \cdot (d_{m_1} - d_{m_2}) \\
   &     d_{m_i} = \sqrt{d^2_{om_i}+d^2_o-2d_{om_i}d \cos \theta_a \cos \theta_e} ~~_{\forall i \in \{1,2\}}
\end{split}
\label{eq:TPD3D} \vspace{-0.1cm}
\end{equation}
where \(\angle\) denotes the phase of a complex number, \(f_s\) is the sampling rate, \(c\) is the speed of sound, and \(F\) is the total number of frequency bands. \(\theta_a\) and \(\theta_e\) represent the azimuth and elevation angles, respectively. \(d_o\), \(d_{m_1}\), and \(d_{m_2}\) are the distances between the target speaker and the microphone(or camera), and from the \(m_1\)-th and \(m_2\)-th microphone to the target speaker, respectively. These distances need to be measured using additional visual devices, such as a depth camera.

In \cite{shao2023rir}, Shao et al. propose interpreting the 3D spatial feature (3D-SF) from an alternative perspective of a multiplicative transfer function (MTF) approximation, as described in \cite{avargel2007multiplicative}. When the STFT of the room impulse response (RIR) from the target source's position to the microphone array is available, denoted as \(R \in \mathbb{C}^{K \times F \times M}\), where \(K\) represents the total length of RIR considered, TPD can also be formulated as:
\begin{equation}
    \text{TPD}_{t,f,(p)} = \angle R_{0, f, m_1} - \angle R_{0, f, m_2}
\end{equation}
where \(\angle R_{0, f, m}\) denotes the phase of the direct wave's RIR from the target source to microphone \(m\).

Intuitively, the more similar the IPD and TPD are, the higher the likelihood that the mixed time-frequency (T-F) bin is dominated by the target source. By computing the cosine of the difference between IPD and TPD, the 3D spatial feature (3D-SF) can be obtained as:
\begin{equation}
    \text{3D-SF}_{t,f} = \cos(\text{IPD}_{t,f,(p)} - \text{TPD}_{t,f,(p)})
\end{equation}

\subsection{RIR Spatial Feature}
\vspace{-2pt}
RIR-SF, introduced in \cite{shao2023rir}, extends the 3D-SF concept by accounting for the impact of reverberant waves on phase differences. It mitigates this effect by convolving the multi-channel speech signal \(Y \in \mathbb{C}^{T \times F \times M}\) with the complex conjugate (H) of the target RIR, denoted as \(R^H \in \mathbb{C}^{K \times F \times M}\), across the time index. This process defines an intermediate phase known as the RIR-convolved phase (RP):
\begin{align}
    \text{RP}_{t,f,m} &= \angle \left( Y * R^H \right)_{t,f,m} \\
    &= \angle \left( \sum_{k=0}^{K-1} Y_{t-k, f, m} \cdot R^H_{k,f,m} \right)
\end{align}
where \(K\) denotes the length of the RIR considered. In alignment with \cite{shao2023rir}, \(K\) is set to 10 through this work, corresponding to a duration of 0.1 seconds, considering a shift size of 10 ms in the STFT. This practice will be adopted in our work to ensure consistency and comparability with established methodologies.

If \(Y_{t,f,m}\) is dominated by the target source, its phase pattern should align with \(R_{t,f,m}\), making \(\text{RP}_{t,f,m}\) independent of the microphone and position. The difference across channels should then approach zero. Analogous to the definition of 3D-SF, RIR-SF is quantified as the cosine similarity of the interchannel RIR-convolved phase differences:
\begin{equation}
    \text{RIR-SF}_{t,f} = \cos \left( \text{RP}_{t,f,m_1} - \text{RP}_{t,f,m_2} \right)
\end{equation}

\section{Proposed: Solo Spatial Feature}

\subsection{Convolving with Solo Segment Instead of RIR}
\vspace{-0.2cm}
While RIR-SF significantly advances spatial feature extraction by considering reverberant effects, its practicality is limited by the difficulty in accessing accurate ground truth RIRs, making it less effective in unpredictable acoustic environments. Acknowledging the practical challenges of acquiring accurate ground truth RIR, our study proposes a novel solution: the Solo Spatial Feature (Solo-SF). By convolving a selected segment from the target speaker's solo part with the mixed speech signal, Solo-SF sidesteps the reliance on precise RIR data. 

To ensure the efficacy of the Solo-SF method, it is imperative that the solo segment \(S \in \mathbb{C}^{K \times F \times M}\) used matches the acoustic environment of the longer speech signal \(Y \in \mathbb{C}^{T \times F \times M}\), specifically \textbf{sharing the same underlying RIR}. It can be garanteed by assuming the target speaker doesn't change it's position during conversation. This alignment enables the effective cancellation of phase patterns when \(S^H\), the conjugate of \(S\), is convolved with \(Y\), mirroring the process employed in RIR-SF. We introduce an intermediate phase termed the Solo-convolved phase (SP), defined as:
\begin{align}
    \text{SP}_{t,f,m} &= \angle \left( Y * S^H \right)_{t,f,m} \\
    &= \angle \left( \sum_{k=0}^{K-1} Y_{t-k, f, m} \cdot S^H_{k,f,m} \right) \label{eq:sp}
\end{align}
Subsequently, the Solo-SF is derived as the cosine similarity between interchannel SP differences:
\begin{equation}
    \text{Solo-SF}_{t,f} = \cos \left( \text{SP}_{t,f,m_1} - \text{SP}_{t,f,m_2} \right)
\end{equation}
This approach not only mitigates the challenges associated with direct RIR measurements but also capitalizes on the inherent vocal traits of the speaker, facilitating a nuanced and robust spatial feature extraction methodology.

\vspace{-0.2cm}
\subsection{Microphone Array Topology and Position Information Independence}
\vspace{-0.2cm}
A significant advantage of the proposed Solo-SF method is its independence from microphone array topology and external position information, such as that obtained from vision-based systems, for determining the target speaker's 3D location or estimating the RIR. 

This independence marks a departure from previous methodologies that required high-quality, audio-visual aligned data, broadening the applicability of models trained with Solo-SF. Consequently, it allows for seamless adaptation to various microphone array configurations without the need for recalibrating or redesigning the spatial feature extraction process based on specific array geometries or visual input.

On the other hand, obtaining the solo part of the target speaker is notably more feasible in practical applications. This can be achieved through activation by wake words or obtained from on-the-fly or offline diarization systems. This accessibility simplifies the process of acquiring clean solo segments, crucial for the effective application of the Solo-SF method in various real-world scenarios.

\begin{figure}
    \centering
    \includegraphics[width=80mm]{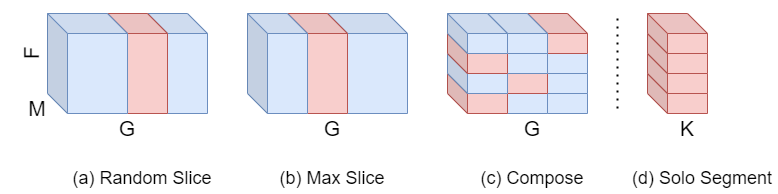}
    \caption{An illustration of 3 different ways of selecting solo segment $S \in \mathbb{C}^{K \times F \times M}$ from a solo part $P \in \mathbb{C}^{G \times F \times M}$}.
    \label{fig:sf}
\end{figure}
\subsection{How To Select Solo Segment?}
\vspace{-0.2cm}
There is a notable special case in Equation \ref{eq:sp} when \(||S^H_{k,f,m}|| \approx 0\), indicating that the target speaker is not producing sound at a particular frequency bin. In such instances, \(S^H_{k,f,m}\) is predominantly influenced by environmental or systematic noise, rendering the solo segment ineffective for distinguishing the desired phase pattern from the mixed speech signal \(Y\). To tackle this issue, we have devised three methods for selecting \(S \in \mathbb{C}^{K \times F \times M}\) from the target speaker's solo part \(P \in \mathbb{C}^{G \times F \times M}\). These methods are designed to ensure that the solo segment remains discriminative and useful for phase pattern extraction, even in challenging noise conditions. The approaches and their implementation are illustrated in Figure \ref{fig:sf}.

\begin{enumerate}

    \item \textbf{Random Slice:} To extract a continuous segment of $k$ frames from the solo part, we employ a random slicing approach as follows:
\begin{equation}
S_{t, f, m} = P_{c+t, f, m},  \quad  c = \text{Rand}[0, G-K]
\end{equation}

\item \textbf{Max Slice:} Select a segment starting from the frame that exhibits the maximum magnitude summation across all frequencies. This process is described as:
\begin{equation}
\label{eq:compose}
    \begin{aligned}
        S_{t, f, m} = P_{c+t, f, m}, \quad c = \underset{t' \in [0, G-K]}{\argmax} \sum_f \left| P_{t', f, m} \right|
    \end{aligned}
\end{equation}

\item \textbf{Compose:} In line with Equation \ref{eq:sp}, since convolution operates on a per-frequency basis, this allows for individual selection of segments for each frequency \(f\), which are then composited into a new segment. Differing from extracting a uniform segment from the solo part, this method selects \(k\) continuous frames with maximum energy for each frequency \(f\) independently, assembling these into the ultimate solo segment:
\begin{equation}
\begin{aligned}
     &S_{t, f, m} = P_{c_f+t, f, m}, \quad c_f = \underset{t' \in [0, G-K]}{\argmax} \left| P_{t', f, m} \right|
\end{aligned}
\end{equation}

\end{enumerate}

\section{Implementation}
\begin{figure}[t]
    \centering
    \includegraphics[width=80mm]{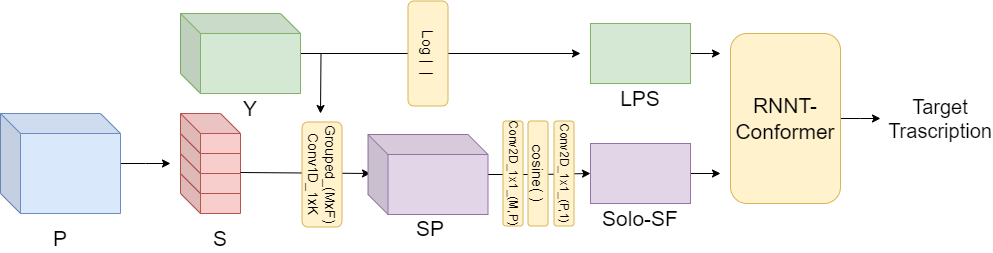}
    \caption{Paradigm of utilizing the target speaker's solo segment \(S \in \mathbb{C}^{K \times F \times M}\), selected from a longer solo part \(P \in \mathbb{C}^{G \times F \times M}\), for target speech recognition in multi-channel, multi-speaker audio \(Y \in \mathbb{C}^{T \times F \times M}\).}
    \label{fig:solo_system}
\end{figure}
Figure \ref{fig:solo_system} illustrates the comprehensive framework we have developed for the extraction of Solo-SF and its application within a speech recognition system. This process initiates with the Short-Time Fourier Transform (STFT) of both the mixed speech signal \(Y\) and the target speaker's solo part \(P\), employing a 25 ms window and a 10 ms hop length at a sampling rate of 16 kHz. All components are efficiently implemented as fully differentiable modules within the PyTorch framework, specifically using \texttt{nn.Module}.

The conversion of Equation \ref{eq:sp} into a computational operation is achieved through the use of \texttt{nn.functional.conv1d}, configured with \(F \times M\) groups, and utilizing the target solo segment \(S \in \mathbb{C}^{K \times F \times M}\) as the convolution kernel. To transition from SP to Solo-SF, two \texttt{nn.Conv2d} layers with specially designed parameters (i.e. 1's and -1's) are employed to compute pairwise interchannel SP differences and their summation, yielding the final Solo-SF. These layers are maintained in a fixed state (frozen) within this study to facilitate future investigations.

Additionally, we extract the logarithmic power spectrum (LPS) from the reference channel, defined as \(\text{LPS}_{t,f} = \log|Y_{t,f,m=1}|\), which serves as a spectral feature. This is then concatenated with the phase-sensitive Solo-Spatial Feature to form a composite feature vector with dimensions \([T \times 2F]\). This composite feature is input into a downstream Conformer-based RNN-T ASR model, as detailed in \cite{kuang2022pruned}, comprising a 12-layer, 4-head Conformer encoder \cite{gulati2020conformer} with 512 attention dimensions and 2048 feed-forward dimensions.

\section{Experiments}
\subsection{Simulated Data: AISHELL-1}
\vspace{-0.1cm}
To ensure our proposed method's comparability with existing approaches, we adopt the data simulation practice outlined in \cite{shao2022multi, shao2023rir} using the AISHELL-1 dataset \cite{bu2017aishell} for our experiments. The Pyroomacoustics toolkit \cite{scheibler2018pyroomacoustics}, leveraging the image-source method (ISM) \cite{lehmann2008prediction}, serves as the basis for Room Impulse Response (RIR) generation and estimation. The simulation parameters include room size, RT60 (reverberation time), microphone position, and speaker positions. Room dimensions vary between \([3, 3, 2.5]\) and \([8, 6, 4]\) meters, featuring one microphone array and two speakers in each scenario. The microphone array consists of an 8-element non-uniform linear array with spacings of 15-10-5-20-5-10-15 cm. RT60 values are randomly chosen from 2 settings, ranging from \([0.1, 0.6]\) seconds and \([0.5, 0.7]\) seconds, reflecting typical room conditions of weak reverberation and strong reverberation respectively. Accordingly, 100,000 sets of RIRs are pre-generated for each configuration. During training, multi-channel reverberant overlapped speech signals are synthesized on-the-fly by convolving the pre-generated RIRs with dry, clean speech samples from AISHELL-1, with signal-to-interference ratios (SIRs) randomly selected between \(-6\) and \(6\) dB. The overlap ratio for the two speakers within a mixed utterance varies from 0.5 to 1, ensuring a diverse set of scenarios for model training and evaluation.

\begin{table}[t]
\centering
\caption{Character Error Rate (CER) \% on AISHELL-1 simulated dev/test sets. \textdagger~ Ground truth relative positions of microphones and target speakers, with uncertainties of $\pm$ 0.5m in their absolute positions within the room. It provides 3D-SF with ground truth $(\theta_a,\theta_e,d_o)$ and provides RIR-SF with partial information for estimating RIR using image source method (ISM). \textdaggerdbl~Indicates the use of Ground Truth RIR (RIR-GT).}
\label{tab:simulate}
\begin{tabular}{lcc}
\hline
\textbf{Method} & \multicolumn{2}{c}{\textbf{CER\% (dev/test)}} \\
\cline{2-3}
 & RT60=(0.1, 0.6)s & RT60=(0.5, 0.7)s \\ \hline
Single-speaker & 8.57/9.71 & 10.64/11.99 \\
3D-GT \textdagger & 14.11/15.67 & 20.28/21.26 \\
RIR-EST \textdagger & 22.37/24.33 & 22.49/24.76 \\
RIR-GT \textdaggerdbl & \textbf{10.03/11.28} & \textbf{10.90/12.38} \\
Solo-random & 16.05/17.99 & 21.39/23.28 \\
Solo-max & 12.70/14.48 & 16.80/18.26 \\
Solo-compose & \textbf{11.57/13.23} & \textbf{13.70/15.56} \\ \hline
\end{tabular}
\end{table}

\begin{table*}[t]
\centering
\caption{Character Error Rate (CER) \% on Alimeeting. \textdagger~Eval-Rev represents evaluation under reverberation without additional interfering speech. \textdaggerdbl~Eval-Simu corresponds to simulation same as that for AISHELL-1 with RT60=(0.1, 0.6)s.}
\label{tab:model_comparison}
\resizebox{\linewidth}{!}{\begin{tabular}{l|l|l|c|c|c|c|c}
\hline
\multirow{2}{*}{\textbf{Method}} & \multirow{2}{*}{\textbf{Training Data}} & \multirow{2}{*}{\textbf{Additional Input}} & \multicolumn{2}{c|}{\textbf{Single-Speaker}} &\multicolumn{3}{c}{\textbf{Multi-Speaker}}\\
& & & \textbf{Eval-Near} & \textbf{Eval-Rev}\textdagger & \textbf{Eval-Simu}\textdaggerdbl & \textbf{Eval-Far} & \textbf{Test-Far} \\ \hline \hline
\multirow{3}{*}{Single-Channel} & Near & \multirow{3}{*}{ None} & 14.72 & 29.02 & 98.08 & 55.25 & 56.28 \\
& Far &  & 20.16 & 23.03 & 87.27 & 34.23 & 37.00 \\
 & Near-Rev\textdagger+Far & & 15.82 & 17.63 & 83.56 & 32.13 & 34.55 \\ \hline
3D-GT & \multirow{5}{*}{Simu\textdaggerdbl} & \begin{tabular}[c]{@{}l@{}}Microphone topology, \\ spk-to-mic relative position\end{tabular} & \multirow{6}{*}{N/A} & 18.96 & 22.72 & \multicolumn{2}{c}{\multirow{3}{*}{N/A}} \\ \cline{1-1} \cline{3-3} \cline{5-6}
RIR-EST &  & \begin{tabular}[c]{@{}l@{}}Estimated RIR (from mic topo, \\ spk and mic absolute position, room size)\end{tabular} &  & 20.94 & 30.04 &\multicolumn{2}{c}{} \\ \cline{1-1} \cline{3-3} \cline{5-6}
RIR-GT & & Ground truth RIR &  & 16.46 & 18.53 & \multicolumn{2}{c}{} \\ \cline{1-3} \cline{5-8}
\multirow{3}{*}{\shortstack{Solo-compose \\ (proposed)}} & Simu\textdaggerdbl & \multirow{3}{*}{2 seconds nearest solo part from diarization} &  & 17.11 & 20.32 & 40.75 & 42.08 \\
 & Far &  &  & 24.79 & 52.23 & 29.59 & 32.12 \\
 & Simu\textdaggerdbl+Far & &  & \textbf{17.59} & \textbf{21.08} & \textbf{26.83} & \textbf{29.52} \\ \hline
 SOT \cite{kanda2020serialized,yu2022m2met} & Near & None & \multicolumn{3}{c|}{\multirow{2}{*}{N/A}} &30.80 &32.40 \\
 $\text{SOT}_{bf}$ \cite{kanda2020serialized,yu2022m2met} &Near+ Far & CDDMA beamformer \cite{huang2020differential} fixed mic topo & \multicolumn{3}{c|}{} &29.70 &30.90 \\ \hline
\end{tabular}}
\end{table*}

\noindent \textbf{Best Solo Segment Selection Method -- Compose:} For an overlapped utterance \(Y\), the solo part \(P\) is consistently selected as a fixed 2-second long, random, continuous speech from the target before mixing with interfering speech. As illustrated in the last three rows of Table \ref{tab:simulate}, employing the "Compose" (i.e. Equation \ref{eq:compose}) method for solo segment selection consistently yields the best results across all three tested methods. This outcome supports our initial hypothesis that maximizing coverage of frequency bins within the solo segment is crucial for optimal performance during convolution.

\noindent \textbf{3D \textit{V.S.} RIR \textit{V.S.} Solo:} RIR-SF, when utilizing ground truth RIR, approaches the performance of the single-speaker lower bound, highlighting its potential efficacy. However, its performance significantly relies on the accuracy of the RIR information. Even when combining partial correct information with ground truth for 3D-SF, its effectiveness diminishes, resulting in outcomes inferior to those achieved by 3D-SF. Conversely, Solo-SF, which requires only a 2-second snippet of solo speech from the target speaker and no vision-based information, delivers results nearly on par with the ideal yet impractical lower bound set by ground truth RIR-SF. Furthermore, it outperforms 3D-SF, establishing Solo-SF as a more dependable alternative in this comparative analysis.

\subsection{Real Data: AliMeeting}
\vspace{-0.2cm}
As discussed in Section 3.2, a distinct advantage of the proposed Solo-SF is its independence from both vision-based inputs and the topology of the microphone array. This attribute significantly enhances its applicability to real-world scenarios and publicly available datasets. For the purpose of our evaluation, we chose the AliMeeting dataset \cite{yu2022m2met}, a Mandarin-language corpus collected from real meeting environments and specifically designed for the ICASSP 2022 M2MeT challenge. This dataset includes multi-channel far-field speech recordings, amassing a total of 104.75 hours for training, 4 hours for evaluation, and 10 hours for testing, alongside simultaneously recorded near-field data from the meeting participants. Each recorded session, ranging from 15 to 30 minutes, features discussions among 2-4 participants. The training data presents an overlap ratio of 42.27\%, which is significantly lower than that of our previously discussed simulated dataset, yet it offers a closer approximation to authentic meeting situations. Furthermore, the challenge documentation stipulates that participants were required to remain stationary during recordings, a condition that dovetails with the operational prerequisites for the effective application of Solo-SF, thereby ensuring the consistency of the Room Impulse Response (RIR) between the solo segments and the speech intended for transcription.

Given the availability of ground truth diarization data from the M2MeT challenge, extracting the solo parts of the target speaker for analysis becomes straightforward. To ensure the consistency of the underlying RIR between the solo segments and the speech to be transcribed, we meticulously select a 2-second solo segment closest in time to the current utterance. This selection is made irrespective of whether the solo part falls within or outside the current utterance, ensuring optimal relevance and acoustic similarity for Solo-SF application.

In alignment with common practices for leveraging open-source data, incorporating near-field clean speech or its simulations into the training data has been shown to enhance ASR model performance. Adhering to this approach, we enriched our training dataset with two variations using the near-field Train set: one with only reverberation (Near-Rev) and another with both reverberation and interference speech (Simu). Given the unavailability of additional inputs for 3D-SF and RIR-SF within the Alimeeting dataset, we generated two evaluation sets from the near-field Eval set, termed Eval-Rev and Eval-Simu, facilitating a direct comparison of Solo-SF with 3D-SF and RIR-SF. As evidenced by the results in Table \ref{tab:model_comparison}, and consistent with observations from AISHELL-1, Solo-SF outperforms its counterparts by demonstrating superior overall performance and robustness across all evaluated spatial features.

Focusing on the real test scenarios, Eval-Far and Test-Far, Solo-SF, when trained on a combination of simulated and far-field data, significantly outperforms the single-channel baseline system—achieving a substantial reduction in CER by absolute margins of 5.30\% and 5.03\%, respectively. Additionally, when compared to the official baseline SOT and its beamformer-enhanced variant, Solo-SF exhibits superior performance without necessitating additional modules. This underscores its efficacy and potential applicability in multi-channel, multi-speaker speech recognition tasks, demonstrating its capability to deliver enhanced speech recognition accuracy in complex acoustic environments.

\section{Conclusion}

In this work, we have advanced the field of multi-channel, multi-speaker ASR by presenting a comprehensive study on the utilization of a target speaker's solo segment. Central to our approach is the introduction of Solo-SF, a novel spatial feature extraction method designed to enhance ASR performance without relying on traditional inputs such as microphone array topology or vision-based data. Our investigation further delves into optimal strategies for selecting solo segments, a critical component for ensuring the effectiveness of Solo-SF in diverse acoustic environments.

Our findings highlight the efficacy of using solo segments for enhancing ASR systems, suggesting avenues for future research in optimizing segment selection. The Solo-SF approach sets a promising direction for improving speech recognition accuracy and robustness in real-world environments.


\bibliographystyle{IEEEtran}
\bibliography{mybib}

\end{document}